\documentclass[12pt]{article}
\usepackage{graphicx} 
\begin{document}
\title{Towards the Unified Theory of Galactic Bar-Modes}
\author{E.V.\,Polyachenko, V.L.\,Polyachenko\\
\itshape Institute of Astronomy RAS, Moscow 109017, 48 Pyatnitskaya st.}
\date{}
\maketitle

\begin{abstract}
The arguments in favor of the unified formation mechanism
 for both slow (Lynden-Bell's) bars and common fast bars are
given. This mechanism consists in a certain instability that is akin
to the well-known radial orbit instability; it is caused by the mutual
attraction and alignment of axes of precessing star orbits (up to now,
such a way of formation was considered only for slow bars).  

The general theory of the low-frequency modes of a disk consisting of
precessing orbits (at different angular velocities) is presented. The
problem of determining these modes is reduced to the integral
equations of a rather simple structure. The characteristic pattern
speeds ($\Omega_p$) of the low-frequency modes are of order of the
mean orbit precession speeds ($\bar\Omega_{pr}$). The bar-modes also
belong to this type of modes. The slow bars have $\Omega_p \approx
\bar\Omega_{pr}$; for the fast bars, $\Omega_p$ may far exceed even
the maximum precessing speed of disk orbits (however, $\Omega_p$
remains to be of order of these precessing speeds). The possibility of
such an excess of $\Omega_p$ over $\Omega_{pr}^{\max}$ is connected
with the effect of ``repelling'' orbits that tend to move in the
direction opposite to that they are being pushed. 

The preliminary analysis of the orbit precession patterns for a number of typical
potentials is given. It is noted that the maximum radius of the
``attracting'' circular orbits ($r_c$) may be used as a reasonable estimate
of a bar length. 
\end{abstract}

\section{Introduction}
It is commonly supposed that a galactic bar can belong to one of two
types --- either common
fast bars or Lynden-Bell slow bars (see, e.g., Sellwood 1993,
Polyachenko 1994). The distinction between them is drawn along a
number of lines.
Firstly, bars with different rotation velocities have considerably
different sizes: while the common bars end up at the corotation
or 4:1 resonance, the Lynden-Bell bars end up in the
vicinity of the inner Lindblad resonance. Secondly, it is believed
that these bars are produced by entirely different formation
mechanisms. For the slow Lynden-Bell bars, the physical mechanism
is absolutely clear~--- this is the mutual attraction and ``sticking''
together of the slowly precessing orbits. Note, that in application to
galactic bars, the idea of axis alignment was first suggested by
Lynden-Bell (1979). 

Yet for the fast bars the situation remains vague and debatable. 
The fast bars were long thought to be generated due to the fast rotation
in much the same way as in the classical incompressible Maclaurin
spheroids being they are strongly oblate and rapidly rotating. It was
shown by Toomre (1981) that in reality the galactic bar-modes have little
in common with incompressible edge modes. Accordingly, they should
have another formation mechanism. Toomre (1981) offered the so-called swing
amplification mechanism, which currently is generally accepted for the 
explanation of the normal SA spirals. We think, however, that a simple
extension of the Toomre mechanism to the SB galaxies can hardly be
done. In particular, it is difficult to suspect the existence of the
running spiral waves, necessary for the swing amplification, which
takes place outside the bar near the corotation. The specific attempt
made by Athanassoula and Sellwood (1986) to demonstrate the validity
of the swing amplification mechanism on some some models of stellar
disks appears to be somewhat artificial.  

In our opinion, it is more preferably to find such a mechanism of the
common bar formation, which is  directly linked to the instability of
the central part of the galactic disk itself.

Contopoulos (1975, 1977) has pointed out that the properties of the
families of the periodic orbits in the rotating bar potentials play
the fundamental role in the theory of the galactic bars. This is
especially true for
the so-called $x_1$ family of orbits, which is elongated along the bar
inside the corotation circle. It is plausible to assume that these
periodic and close non-periodic orbits make up the galactic bar. One
should note, however, that the preceding is relevant to the theory of
the already existing bars, and in the strict sense cannot be applied
the the bar formation mechanism itself\footnote{This is especially true
  for the linear stage of the bar-forming instability (if the
  bar arose due to the growth of the instability): the orbits
  considered by Contopoulos are obviously trapped by the potential of
the bar. The trapping process is, of course, the non-linear
phenomenon.}. 

Nevertheless, from the described picture of orbits constituting 
the bar, it is customary to make a ``reasonable'' conclusions about
the mechanisms of bar formation, including the possible instability at
the linear stage. The fast bar angular velocity $\Omega_p$ is greater
(sometimes substantially greater) than the maximum precession velocity
$\Omega_{pr}^{\max}$. Meanwhile, it is intuitively suggested that the angular
modulation of the precessing orbit distribution (i.e., a~figure of the bar)
should rotate with the velocity of about mean precession velocity
of the orbits. Such a conclusion would have meant the uselessness of
the Lynden-Bell mechanism for describing the fast bar formation. So,
it is generally agreed that in reality the growing bar forces the
orbits to change their shape and tune to the strengthening and narrowing
bar. Moreover, it is assumed that this effect can be significant even
at the linear stage. It is justified on the example of 
weak bars in the framework of the linear theory (Sellwood
1993): the theory shows that initially circular orbits change into
slightly elongated ovals, with orientation relative to the bar just as
the orbits of the $x_1$-family. 

However, one can adduce two important objections as to the said in
the last paragraph.

1. The circular orbits cannot be the typical sample orbits of the
   undisturbed disk (excepting for a cold disk model, which is of little
   interest in this case) until the bar-induced perturbed velocities
   do not exceed the velocity dispersion in the original axisymmetric
   disk. It is clear that this condition should be met first at the
   initial stage of the instability development.

2. The intuitive conclusion mentioned above (that the velocity of the
   wave density in the system of precessing orbits, $\Omega_p$, cannot
   exceed the maximum precession velocity    $\Omega_{pr}^{\max}$) is
   supported by the results of computations of the unstable modes 
   in numerous models of galactic disks. These computations
   show that in all cases the pattern speed $\Omega_p$ is smaller than
   the maximum star angular velocity $\Omega_{\max}$ in the
   disk\footnote{Note that this property of the oscillation
   frequency spectrum of the gravitating disk is not yet found for the
   general case.}. But this analogy appears to be somewhat loosely after
   the discovering of the possible ``donkey'' behavior of the star orbits
   by Lynden-Bell and Kalnajs (1972), when the orbit accelerates when
   held back and slows down when urged forward.
   Such a behavior takes place under the Lynden-Bell condition,
   $\left.\partial \Omega_{pr}/\partial L\right|_{J_f} < 0$
   (Lynden-Bell 1979), where $L$ is the star angular momentum, $J_f =
   J_r + L/2$ is the Lynden-Bell adiabatic invariant , $J_r$ is the
   radial action. As it is shown below, if the orbits with a
   ``donkey'' behavior play an active role in the bar-forming
   instability, the bar angular velocity can exceed $\Omega_{pr}^{\max}$.

The last argument allows to treat the fast bar-modes as the
corresponding density waves of the precessing orbits, in the
full correspondence with the theory of the Lynden-Bell bars. Note that
Kalnajs (1973) was first who called attention to the  possibility
of the freely precessing orbit alignment (and thus the bar formation). The
important point in his theory of kinematic waves in the case of
near-circular orbits was the independence of the precession velocity 
$\Omega_{pr}(r) = \Omega(r) - \kappa(r)/2$ ($\Omega(r)$ from radius
(here $\kappa(r)$ is the epicyclic frequency, $\kappa^2 = 4\Omega^2 +
rd\Omega^2/dr$). Note that the same condition $\Omega_{pr}(r)
\approx \mathrm{const}$ is required in the Lynden-Bell (1979) more
general consideration of the problem. 

In reality different stars precessing with different velocities. But only
important point is that the angular precession velocities (and also the
pattern speeds of bars ($\Omega_p$) in the majority of cases) are substantially
smaller than the typical star azimuthal frequencies $\Omega_2$ and especially
the radial frequencies $\Omega_1$. If for some orbit the inequality
\begin{equation}
|\Omega_p - \Omega_{pr}|/\Omega_1 \ll 1 
\label{cond1}
\end{equation}
holds, this whole orbit (not individual stars) can be considered as an
elementary object in the interaction with the bar gravitational
field. If the condition (\ref{cond1}) holds for the majority of disk
orbits, participating in the bar formation, we can consider the
processes (e.g., bar-instability) in the model disk, consisting of the
set of the precessing stars. As it is noted by Lynden-Bell
(1979), the condition (\ref{cond1}) means the conservation of the
adiabatic invariant $J_f = J_r + L/2$, that reduces the problem of
the bar-instability to the one dimensional problem: one have to follow
only the variation of the azimuthal positions of the orbit major axes
under the action of the gravitational attraction from the bar. 

Actually, we solve the problem of bar density wave of orbits with
different precession velocities, that is quite analogous to more
familiar problem of the star density waves (e.g., of the bar-like
shape) in the differentially rotating disks. Seemingly, not all the
orbits participating in the bar-mode formation, meet the condition
(\ref{cond1}) with margin, especially for the fast bars. But even for the latter
usually $|\Omega_p - \Omega_{pr}| \sim \Omega_{pr} \sim \omega_G$,
where $\omega_G\sim\sqrt{GM_d/a^3}$ is the characteristic
gravitational (Jeans) frequency ($G$  is the gravitational constant,
$M_d$ is the mass of the active disk, $a$ is its radius). Then when
$\Omega_{pr}/\Omega_1 \ll 1$ we are still within the bounds of the
condition (\ref{cond1}). Even in case of more weak inequality
$\Omega_{pr}/\Omega_1 < 1$  (e.g., several times smaller but not at
some orders) the suggested model will likely provide the correct
qualitative answer. This is at any case not worse than, for example,
the analysis of the spiral structure of the galaxy M\,33 by Shu
{\it et al.} (1971) by using the WKB formulae from the well-known theory
of Lin and Shu, applicable, strictly speaking, to the tightly wound
multi-turn spirals.

A few words about the content of the paper. In the Section 2, we
descibe in the general form a model of the precessing orbits and
derive the basic equations for the model, useful for the analysis of 
the disk low frequency modes of interest for us. The use of 
the action--angle variables $I_1$, $I_2 = L$ and $w_1$, $w_2$ is the
easiest way for derivation of these equations from the general
kinetic equation. The substantial simplification
appears with introducing a slow angular variable $\bar w_2 = w_2 -
w_1/2$ and averaging over the fast angle variable $w_1$ (i.e., over the
radial stellar oscillations). In the simplest cases,  the problem can
be reduced to the analysis of a rather simple dispersion
relations. In some more general cases, we obtain the integral equations
of different degrees of complexity. But even in the most general form,
the obtained integral equations for this model is much simpler that
the immense integral equations for the normal modes of the stellar
disk, obtained by Kalnajs (1965) and Shu (1970). Remind that the
use of these general integral equations (not counting the $N$-body
methods) was the only possibility for analysis of large-scale
modes, as opposed to incomparably simpler problem of tightly-wound
spirals . This is especially true in regard to the bar-mode.
As discussed above, the real simplification has come through the
analysis of low-frequency modes. Note that the resulting description
of gravitating systems is similar to the drift approximation in the
plasma physics (see, e.g., Chew  et. al., 1956), but for orbits of
essentially more general type. In our opinion, the most important  
advantage of our approach consists in the fact that it makes clear the
simple physical mechanisms of the  instability processes developing in
a disk. Unfortunately, a possibility of revealing these physical
mechanisms under the use of the general integral equations by Kalnajs
or Shu would be practically impossible, and the same is true for the
$N$-body simulations. 
 
In Sec. 3, we analyze the dispersion relation for a model disk
consisting of orbits of two different types  that differ by their
precessing speeds ($\Omega_{pr}^{(1)}$ and $\Omega_{pr}^{(2)}$;
$\Omega_{pr}^{(2)} > \Omega_{pr}^{(1)}$) and, generally speaking,
also by a sign of the Lynden-Bell derivative  $(\partial
\Omega_{pr}/\partial L)_{J_f} \equiv \Omega'_{pr}$ .For the case when 
the derivatives $(\Omega'_{pr})^{(1)}$ and $(\Omega'_{pr})^{(2)}$
have opposite signs (i.e., a disk contains both ``attracting'' and
``repelling'' orbits), the pattern speed of the unstable modes
${\mathrm Re}\, \Omega_p$ may be more than $(\Omega_{pr})^{(2)}$,
i.e. the maximum orbit precessing speed in a disk. We show that at the
same time $\Omega_{pr}^{(1)} < \mathrm{Re}\,\Omega_p <
\Omega_{pr}^{(1)}$ when $(\Omega'_{pr})^{(1)}$ and
$(\Omega'_{pr})^{(2)}$ have the same sign  (positive for
instability). We consider this result as the important argument in
favor of the unified formation mechanism both for slow and fast bars.

In Sec. 4, the results of computation of the precessing speeds
$\Omega_{pr}$ and the most important for the theory quantity, the
Lynden-Bell derivative $\Omega'_{pr}$, are given for a number of
typical potentials. In particular,  the regions on the Lynden-Bell
$J_f, L$ plane where $\Omega'_{pr} > 0$ and $\Omega'_{pr} < 0$ are
found. The natural suggestion is made that a bar forms by
``attracting'' orbits with $\Omega'_{pr} > 0$\footnote{Note that in
the formation process for all the mode, i.e. both the bar and adjacent 
spirals, and not just the central bar, the ``repelling'' orbits with
$\Omega'_{pr} < 0$ may also take part. Moreover, this may be important
to explain the fast bar phenomenon (as discussed above).}. Then the
bar length (more exactly, the radius of the bar's end) should be equal  
to the maximum of the apogee radii ($r_{\max}$), among the
sufficiently occupied  orbits from the region where $\Omega'_{pr} >
0$. So, for calculation of $l_b$, one should know not only the general
pattern of orbit precessions determined by  the potential $\Phi_0(r)$
but a specific equilibrium distribution function as well. As a first
approximation, one can take the estimate $l_b \sim r_c$, where $r_c$
is the radius of the circular orbit at which $\Omega'_{pr} = 0$
($\Omega'_{pr} > 0$  for $r < r_c$, and $\Omega'_{pr} < 0$ for $r >
r_c$, at the circular orbits).  These radii $r_c$ are computed for all
the potentials considered. Note that our  determination of the bar
length has nothing to do with common determinations that link this
length with a location of one of the resonances (CR, ILR or 4:1). It
is clear  that our bar length may take a great variety of values
depending on the specific  potential and distribution function
(accidentally, they may fall near one of resonances). 

In Sec. 5, we shortly formulate the most important conclusions and
discuss some immediate prospects for a work in this field.

\section{Basic equations for the low-frequency modes}
For derivation of the basic equations of the theory, it is most
conveniently to use the action-angle variables ${\bf I} = (I_1, I_2)$
and ${\bf w} = (w_l,w_2)$, which suitably takes account of the 
double periodicity of stellar motion in the equilibrium potential. 
Note that $I_1 = I_r$ ($I_r$ is the radial action), $I_2 = L$ ($L$ is
the angular momentum). We start out with the linearized kinetic
equation in its usual form (see, e.g., Fridman and Polyachenko 1984):
\begin{equation}
\frac{\partial f_1}{\partial t} + \Omega_1\frac{\partial f_1}{\partial
  w_1} + \Omega_2\frac{\partial f_1}{\partial w_2} = 
\frac{\partial f_0}{\partial I_1}\frac{\partial\Phi_1}{\partial w_1}+
\frac{\partial f_0}{\partial I_2}\frac{\partial\Phi_1}{\partial w_2},
\label{kint2}
\end{equation}
where $f_0({\bf I})$ and $f_1({\bf I},{\bf w}, t)$ are the unperturbed
and perturbed distribution functions, $\Phi_1$ is the perturbation of
the gravitational potential, $\Omega_1$ and $\Omega_2$ are the
frequencies of the radial and azimuthal oscillations of stars in the
equilibrium potential $\Phi_0(r)$, $\Omega_i = \partial E({\bf
  I})/\partial I_i$ ($E$~is the energy in terms of ${\bf I}$,
$i=1,2$). The change of variables $\bar w_2 = w_2 - w_1/2$, $\bar w_1
= w_1$ in (\ref{kint2}) yields
\begin{equation}
\frac{\partial f}{\partial t} + im\Omega_{pr}f + 
\Omega_1\frac{\partial f}{\partial  w_1} =
\frac{\partial f_0}{\partial I_1}\frac{\partial\Phi}{\partial w_1}+
im\Phi\left(\frac{\partial f_0}{\partial I_2}-\frac12\frac{\partial
  f_0}{\partial I_1}\right), 
\label{kint3}
\end{equation}
where we have assumed that the perturbations are proportional to
$\exp(im\bar w_2)$:
$$
\Phi_1 = \Phi \exp(im\bar w_2),\quad f_1 = f\exp(im\bar w_2),
$$
$m$ is the azimuthal index (an even integer), and $\Omega_{pr}(E,L)
=\Omega_2 - \Omega_1/2$ is the precessing speed of an orbit with 
energy $E$ and angular momentum $L$. If we also transform from
the action ($I_1$, $I_2$) to ($E$, $L$) (the equilibrium distribution
function is usually given just in the variables $E$, $L$), we obtain
the linearized kinetic equation in the form
\begin{equation}
\frac{\partial F}{\partial t} + im\Omega_{pr}F + 
\Omega_1\frac{\partial F}{\partial  w_1} =
\Omega_1\frac{\partial F_0}{\partial E}\frac{\partial\Phi}{\partial w_1}+
im\Phi\left(\frac{\partial F_0}{\partial L}+\Omega_{pr}\frac{\partial
  F_0}{\partial E}\right), 
\label{kint4}
\end{equation}
where $F_0(E,L) = F_0(I_1,I_2)$. Note that in this form,
Eq. (\ref{kint4}) also holds for a quasi-Coulomb potential $\Phi_0$,
i.e., one that is due principally to a large central mass. All that is 
necessary then is to redefine $\bar w_2$ and $\Omega_{pr}$:
${\bar w_2} = w_2 - w_1$, $\Omega_{pr} = \Omega_2 - \Omega_1$.

Then, neglecting self-gravitation of the system and the quadrupole
moment of the central object, we have $\Omega_{pr} = 0$, as it should
be for closed Keplerian orbits. Here the azimuthal index $m$ can be
either odd or even.

As it is shown by Lynden-Bell (1979) (see details in the Introduction)
the most adequate variables that should be used instead ($I_1$, $I_2$)
or ($E$, $L$) are $J_f = I_1 + I_2/2$ and $L = I_2$. Changing in
(\ref{kint3}) to these variables, we obtain the kinetic equation in
the most convenient form for studying the low-frequency modes:
\begin{equation}
\frac{\partial f}{\partial t} + im\Omega_{pr}f + 
\Omega_1\frac{\partial f}{\partial  w_1} =
\frac{\partial {\cal F}_0}{\partial J_f}\frac{\partial\Phi}{\partial w_1}+
im\Phi\frac{\partial {\cal F}_0}{\partial L}, 
\label{kint5}
\end{equation}
where ${\cal F}_0(J_f,L) = f_0(I_1,I_2)$ and taken into account that
$\partial f_0/\partial I_2 - \frac12\partial f_0/\partial I_1 =
  \partial F_0/ \partial L$.

We assume that the rms deviation of the precession rates about the
mean $\bar\Omega_{pr}$, given by $\Delta\Omega_{pr} =
[\overline{(\Omega_{pr} - \bar\Omega_{pr})^2}]^{1/2}$ 
and the typical gravitational frequency $\omega_G$ are both small,
$\bar\Omega_{pr}$,~$\omega_G \ll \Omega_1$. Note that
$\omega_G$ is of the order of the mean Jeans frequency of the system:
$\omega_G \sim \sqrt{GM_d/a^3}$, where $M_d$ and $a$ is the mass and
the radius of the active disk, respectively.
The above inequalities can be justified, e.g., if one assume that we
are dealing with a system of stars, that have close precession speeds,
within a massive halo which, while remaining unpertubed itself, 
furnishes the dominant contribution to the equilibrium potential
$\Phi_0$. However, it should be remembered that the role of halo
should not be played only by the real massive spherical component of 
the galaxy. Indeed, since in general the extent to which  different
groups of orbits are involved into the instability process can differ
greatly, one can consider as the first approximation that the active
group of stars is immersed in the massive halo of other disk stars. 
Under these circumstances, then, there may exist a low-frequency
mode ($\propto\exp(-i\bar\omega t)$, with $\bar\omega\equiv \omega -
m\bar\Omega_{pr}\sim \omega_G$, $\Delta\Omega_{pr}$) in a
coordinate system rotating at angular velocity $\bar\Omega_{pr}$ such
that the slow precessional dispersal of orbits is canceled by their
mutual gravitational attraction. It would be natural to suppose that
if self-gravitation were to win out over the dispersion in orbital
precession speeds, an instability should develop that could eventually
deform the system (under the influence of the largest-scale growing
modes). It is clear, however, that even in a system with essentially
radial orbits, this holds true only if the torque that alters the
orbital angular momentum of the stars forces their orbital precession 
speeds to change in the same direction (see the Introduction and,
e.g., the formula (\ref{k13}) below). 

We use the perturbation theory to derive the desired solution for the
low-frequency modes. Let ${\cal F} = {\cal F}^{(1)} + {\cal F}^{(2)}$,
where ${\cal F}^{(1)}$ corresponds to the permutational mode obtained
from (\ref{kint5}) by neglecting terms proportional to
$\Delta\Omega_{pr}$ and $\Phi\propto G$: $\omega=0$ (or $\omega =
m\bar\Omega_{pr}$ when $\bar\Omega_{pr} \ne 0$), $\partial{\cal
  F}^{(1)}/\partial w_1=0$, i.e. ${\cal F}^{(1)} = {\cal
  F}^{(1)}(J_f,L)$ is an as-yet arbitrary function of the integrals of 
motion, which we will subsequently specify by requiring that the
solution of the next approximation be periodic. The equation for
${\cal F}^{(2)}$ takes the form
\begin{equation}
-im{\cal F}^{(1)} + im\Omega_{pr}{\cal F}^{(1)} + 
\Omega_1\frac{\partial {\cal F}^{(2)}}{\partial  w_1} =
\frac{\partial {\cal F}_0}{\partial J_f}\frac{\partial\Phi}{\partial w_1}+
im\Phi\frac{\partial {\cal F}_0}{\partial L}.
\label{kint6}
\end{equation}
Averaging (\ref{kint6}) over $w_1$, from $0$ to $2\pi$, and bearing in
mind the periodicity of the functions ${\cal F}^{(2)}$ and $\Phi$, we
have 
\begin{equation}
-(\bar\omega - m\delta\Omega_{pr}){\cal F}^{(1)}\approx \frac{\partial
 {\cal F}_0}{\partial L}\frac1{2\pi} \int\limits_0^{2\pi}\Phi dw_1,
\label{kint7}
\end{equation}
where $\delta\Omega_{pr} = \Omega_{pr} - \bar\Omega_{pr}$.

Invoking the Poisson equation, some minor manipulations yield
\begin{equation}
\Phi = -G\int d{\bf J}'{\cal F}^{(1)}({\bf J}')\int d{\bf w}'
\Gamma(r,r',\varphi'-varphi) \exp[im(\bar w'_2 - \bar w_2)],
\label{i8}
\end{equation}
where $d{\bf J}' = dJ_f'dL'$, $d{\bf w}' = dw_1'dw_2'$, $\Gamma$ is the Green
function:
\begin{equation}
\Gamma = \frac1{r_{12}},\quad r_{12} = [r^2+r'^2-2rr'\cos(\varphi'-\varphi)]^{1/2}.
\label{g9}
\end{equation}
Equation (\ref{i8}) is an integral equation for the function
$\Phi({\bf J}, w_1)$,) if one takes into account the expression
(\ref{kint7}) for ${\cal F}^{(1)}$ through $\Phi$. The coordinates of
stars $r$, $\varphi$, $r'$, $\varphi'$ in (\ref{i8}) and (\ref{g9})
must be expressed in terms of ${\bf J}$, ${\bf J}'$, ${\bf w}$, ${\bf
  w}'$, where 
$$
r=r({\bf J},w),~r'=r'({\bf J}, w'_1),~\bar w'_2-\bar w_2 =
(w'_2-w_2)-(w'_1-w_1)/2, 
$$
$$
\varphi'-\varphi\equiv\delta\varphi = w'_2 -w_2 + \phi({\bf J},{\bf J}', w_1,w'_1)
$$
and we refrain from writing out the expression for $\phi$. Since the
perturbed potential can always be written out as
$$
\Phi_1(r,\varphi) = \bar\Phi_1(r)\exp(im\varphi)=\Phi\exp(im\bar w_2),
$$
we have
$$
\Phi(r,\varphi) = \bar\Phi_1(r)\exp[im\delta({\bf J},w_1)],
$$
($\delta = \varphi - \bar w_2$ is a known function of ${\bf J}$ and
$w_1$). In actual fact, then, (\ref{i8}) is an integral equation for
the unknown function $\Phi_1(r)$ of only one variable. This equation
can be rewritten in more symmetric form. The right-hand side of
(\ref{i8}) depends on $w_1$ only through $\Gamma$  and
$\exp[im(\bar w'_2 - \bar w_2)]$. So, by averaging (\ref{i8}) over
$w_1$, we obtain for the function 
$$
\chi({\bf J}) = \bar\Phi = \frac1{2\pi}\int\limits_0^{2\pi}\Phi dw_1 
$$
the following integral equation
\begin{equation}
\chi({\bf J}) = \frac1{2\pi}\int d {\bf J}' \Pi({\bf J}, {\bf J}')
\frac{\partial {\cal F}_0({\bf J}')/\partial L'}{\omega-m\delta
  \Omega_{pr}({\bf J}')} \chi({\bf J}')
\label{i10}
\end{equation}
where
\begin{equation}
 \Pi({\bf J}, {\bf J}') = \int dw_1dw_1'd\delta
 w_2\Gamma(r,r',\delta\varphi) \exp(im\delta w_2)\exp[-im(w'_1-w_1)/2],
\label{i10s}
\end{equation}
and $\delta w_2\equiv w_2' - w_2$.

Physically, $\Pi({\bf J}, {\bf J}')$ is proportional to the torque
$\delta M$ acting upon some selected (test) orbit with the action
${\bf J}$ resulting from all orbits with fixed action ${\bf J}'$;
these all have the same shape, but their major axes are oriented in
all possible directions: 
\begin{equation}
\delta M\propto -imG \exp(im\bar w_2)\Pi({\bf J}, {\bf J}'){\cal
  F}^{(1)} ({\bf J}')d {\bf J}'.
\label{i10ss}
\end{equation}

For a quasi-Coulomb field $\Phi_0(r)$, instead of the Lynden-Bell
integral $J_f = I_1 + I_2/2$ we have to use $J_f^C = I_1 + I_2$ in (\ref{i10})
and instead of $\exp[-im(w'_1 - w_1)/2]$ (with $m$ required
to be even) in (\ref{i10s}) we have to use $\exp[-im(w'_1 - w_1)]$
(with arbitrary $m$).

One might hope to reduce (\ref{i10}) to one-dimensional integral
equations in two limiting cases: 1) when the distribution function
${\cal F}_0({\bf J})$ is close to a delta function in $L$ near some
value $L_0$; we will be commenting on this circumstance, writing
${\cal F}_0 = \Delta_1(J_f, L-L_0)\approx \delta(L-L_0)
\varphi_0(J_f)$; 2) for the systems with near-circular orbits,
whereupon $f_0 = \Delta_2(I_1,I_2)$, where $\Delta_2\approx
\delta(I_1) \bar\varphi_0(I_2)$. Below we will restrict ourselves
mainly with the first case. The second case is technically somewhat
more complicated; we will study this important case elsewhere.

Thus we assume ${\cal F}_0 = \Delta_1(J_f, L-L_0)$. Using the fact
that $\Pi$ and $\chi$ vary in $L'$ only a little over the
characteristic scale length of the function ($\partial{\cal F}_0/
\partial L)/[\bar\omega-m\delta\Omega_{pr}(J_f',L')]$, we can reduce
Eq.(\ref{i10}) to an integral equation for the function of one variable 
$\psi(J)\equiv\chi(J, L' = L_0)$; for brevity, hereafter we omit the
index ``$f$'' in the Lynden-Bell integral $J_f$:
\begin{equation}
\psi(J) = \frac{Gm}{2\pi}\int dJ' P(J,J')S_0(J')\psi(J'),
\label{i11}
\end{equation}
where
\begin{equation}
P(J,J') = \Pi(J,J',L=L_0,L'=L_0),
\label{i11s}
\end{equation}
\begin{equation}
S_0(J) = \int d L' \frac{\partial{\cal F}_0(J',L')/ \partial
  L'}{\bar\omega-m\delta\Omega_{pr}(J',L')}. 
\label{i11ss}
\end{equation}

More convenient form for the function $S_0(J')$ is obtained after
integration (\ref{i11s}) by parts
\begin{equation}
S_0(J') = -m\int dL'\frac{{\cal F}_0(J',L')\partial\Omega_{pr}(J',L')/ \partial
  L'}{[\bar\omega-m\delta\Omega_{pr}(J',L')]^2},
\label{i11sss}
\end{equation}
where the derivative $\partial\Omega_{pr}/\partial L'$ can be taken at
$L' = L_0$.

If we are dealing with near radial orbits, we can put $L_0 = 0$
when we calculate $P(J,J')$. Furthermore,$\delta \varphi \equiv
\varphi' - \varphi \approx \bar w'_2 - \bar w_2$ for such orbits, so
the function $\Pi$ can then be substantially simplified:
\begin{equation}
\Pi({\bf J}, {\bf J}') = \int dw_1 dw_1' J_m[r({\bf J},w_1), r({\bf
    J}',w'_1)], 
\label{t12}
\end{equation}
where
\begin{equation}
J_m(r,r') =
\frac1{2\pi}\int\limits_0^{2\pi}d\alpha\Gamma(r,r',\alpha)\cos m\alpha.
\label{t12s}
\end{equation}

When the orbits are exactly radial (``cold'' system), i.e.,
${\cal F}_0 = \delta(L)\varphi_0(J)$,
\begin{equation}
S_0(J') = -\frac{m}{\bar\omega^2}A(J')\varphi_0(J'),
\label{k13}
\end{equation}
where
\begin{equation}
A(J') = \left.\frac{\partial\Omega_{pr}(J',L')}{\partial L'}\right|_{L'=0}.
\label{k13s}
\end{equation}
Accordingly, the integral equation (\ref{i11}), which then describes
the radial orbit instability in a cold system, is
\begin{equation}
\psi(J) = -\frac{Gm^2}{2\pi\omega^2}\int dJ'P(J,J')A(J')\varphi_0(J')
\psi(J'). 
\label{i14}
\end{equation}

It can easily be shown that $J_m(r,r')$ defined by (\ref{t12s}) is a
positive function. Consequently, so is $\Pi({\bf J}, {\bf J}')$ from
(\ref{t12}), and most importantly, so is $P(J,J')$ from (\ref{i11s})
when $L_0 =0$; just the function $P(J,J')$ enters into the derived
equation (\ref{i14}). This can be made explicitly by expanding the
function $(r^2 + r'^2 - 2rr'\cos\alpha)^{-1/2}$ in a harmonic series,
we obtain 
$$
J_m(r,r') = \sum\limits_{n=0}^\infty F_n(r,r')\frac1\pi\int\limits
_0^\pi d\alpha P_n(\cos\alpha)\cos m\alpha, 
$$
where
\begin{equation}
F_n(r,r') = r_<^n/r_>^{n+1},\quad r_<\equiv\min(r,r'),\quad r_>\equiv
\max(r,r'), 
\label{i15}
\end{equation}
and the $P_n$ are Legendre polynomials. Now the positive definiteness of
$J_m$ follows from the fact that $P_n(\cos\alpha)$ can in turn be expanded in
cosines of multiples of the angle, with positive coefficients
(Gradshtein, Ryzhik 1965):
$$
P_n(\cos\alpha) = \frac{(2n-1)!!}{2^{n-1}n!}\left[\cos n\alpha +
  \frac11 \frac{n}{2n-1}\cos(n-2)\alpha+...\right]\equiv
{\sum\limits_k} 'A_k^{(n)}, \cos k\alpha
$$
with all $A_k^{(n)} > 0$ (a prime indicates that the parity of $k$ and
$n$ must be the same), and from
$$
\frac1\pi\int\limits_0^\pi d\alpha \cos m\alpha \cos n\alpha =
\frac{\delta_{mn}}2. 
$$
We finally obtain a convenient representation for $J_m(r,r')$ in simple
series form:
$$
J_m(r,r') = \frac12{\sum\limits_{n\ge m}}' F_n(r,r')A_m^{(n)} > 0.
$$

Given the positivity of $P(J,J')$, the sign of the integrand in
(\ref{i14}), and therefore the sign of $\omega^2$ (i.e., the stability
or instability of a system with purely radial orbits), will depend on
the sign of $A(J')$ as defined by (\ref{k13}). If $A > 0$ for all
orbits in the system under consideration (in other words, for all
values of $J$ or what amounts to he same thing in the present case,
for any value of the energy $E$ of radial stellar oscillations), then
$\omega^2 < 0$. As a result, radial orbits are unstable with $A >
0$. On the other hand, provided that $A < 0$, it is the purely
oscillatory mode. The most compact formula for computing $A(E)$ is
$$
A(E) = \frac1{(2E)^{1/2}} \frac{\displaystyle
\lim\limits_{r_0\to 0} \left\{
\int\limits_{r_0}^{r_{\max}} \frac{dx}{x^2[1-\Phi_0(x)/E]^{1/2}} -
\frac1{r_0}\right\}}{\displaystyle \int\limits_0^{r_{\max}}
  \frac{dx}{[2E-2\Phi_0(x)]^{1/2}}}  
$$

The inequality $(\partial\Omega_{pr}/\partial L)|_{L=0} >0$ is merely
a necessary (and in no way sufficient) condition for the radial orbit
instability, and in particular, for the formation of a bar. The
insufficiency of this criterion is immediately obvious from the fact
that the retarding torque due to the bar can turn out to be
ineffectual in the face of large orbital precession speeds. To derive
valid conditions for bar formation, it is necessary to solve the
problem of stabilization of the radial orbit instability by some
finite dispersion of the orbit precession speeds [having demonstrated
once again the predominance of the bar mode ($m = 2$)]. The bar
formation criterion is in fact none other than the condition for
the bar-mode instability of the type under consideration. We therefore
now proceed to derive the stabilization conditions for systems with
near-radial orbits. 

For definiteness, we assume the Maxwellian distribution in $L$
\begin{equation}
f_0 = \frac1{\pi^{1/2}L_T} \exp(-L^2/L_T^2)\varphi_0(J),
\label{m16}
\end{equation}
where $L_T$ is the thermal spread. If we then assume in (\ref{i11ss})
that $\bar\omega = \omega = 0$, $L_0 = 0$, $\delta\Omega_{pr} =
\Omega_{pr} \approx A(J)L$, we will then have for the system's stability
boundary 
$$
S_0(J') = 2\varphi_0(J')/mL_2^2A(J')
$$
so that the integral equation (\ref{i11}) becomes
\begin{equation}
\psi(J) = \frac{G}{\pi L_T^2}\int d J' P(J,J')
\frac{\varphi_0(J')}{A(J')}\psi(J'). 
\label{i17}
\end{equation}
This equation is almost the same as Eq. (\ref{i14}). A comparison of
these two equations suggests a simple relationship between the
instability growth rate $\gamma$ for a system with purely radial
orbits, $\gamma^2 = -\omega^2$, and the minimum dispersion in orbital
angular momentum required to supress that instability:
\begin{equation}
(L_T)_{\min} = 2^{1/2}\gamma/m\bar A,
\label{r18}
\end{equation}
where $\bar A$ is some mean over the stellar orbits with different
energies $E$. 

The relation (\ref{r18}) acquires a precisely defined meaning when all
stars have almost the same energy, $E\approx E_0$, since we can then
take $\bar A = A(E_0)$. If in (\ref{m16}) we go to the distribution
over precession speeds, $\Omega_{pr} = AL$, we then obtain a more
obvious relation in place of (\ref{r18}):  
\begin{equation}
(\Omega_{pr})_T = 2^{1/2}\gamma/m,
\label{r19}
\end{equation}
where $(\Omega_{pr})_T$ denotes the thermal spread in precession
speeds, and the growth rate $\gamma$ is given in the form $\gamma(m)$
to emphasize that in general it depends on the azimuthal index
$m$. Since $\gamma(m)$ is only a weak function of $m$\footnote{For
  example, we have $\gamma(m)\propto m^{1/2}$ for $m\gg1$. Then, the
  equation (\ref{t12}) is simplified since $J_m \approx
  2\delta(r-r')/m$ for $m\gg1$. Thus, one of the integrations in
  (\ref{t12}) can be
  performed. The asymptotic expression for $J_m$ is most conveniently
  derived directly from Poisson's equation, assuming that
  $m^2\Phi_1/r^2\gg |d^2\Phi_1/ dr^2|$, $|r^{-1}d\Phi_1 / dr|$ in the
  latter.}, it follows from (\ref{r19}) that the most 
difficult modes to stabilize (and in that sense, the most unstable)
are those with the smallest possible $m$. For almost radial orbits, we
have $m_{\min} = 2$, which corresponds precisely to formation of an
elliptical bar out of an initially circular disk. All the modes with
odd $m$, particularly the $m = 1$ mode, are suppressed in this case,
as two oppositely directed (but equal) moments of forces would act on
the two halves of an elongated orbit. The forces break, but do not
rotate such ``needle'' orbits. 

For near-circular orbits in a potential close to that produced by
a central point mass, however, the $m = 1$ mode is immediately become
dominating.

In conclusion to this Section, we give the integral equation of a type
(\ref{i11}) in a form, suitable for computation of the eigen
frequencies of the low-frequency modes of a stellar disk with the
equilibrium distribution function $f_0(E,L)$, provided that a disk
contains substantial fraction of elongated orbits:
\begin{equation}
f(E_1) = \int\limits_{E_{\min}}^{E_{\max}} K(E_1,E_2)f(E_2)dE_2,
\label{i20}
\end{equation}
where the kernel is equal to 
\begin{equation}
K(E_1,E_2) = -\frac\pi{M_1(E_1)} \int\limits_0^{L_{\max}}
\frac{f_0(E_1,L_1)}{(\Omega_p -\Omega_{pr}^{(1)})^2}
\Omega'_{pr}(E_1,L_1) dL_1 \int\limits_0^a\int\limits_0^a dx\,dy\,
\rho^{(E_1)}(x) \rho^{(E_2)}(y) J_m(x,y),
\label{k21}
\end{equation}
Such a form of the integral equation suggests that the torque of
forces produced by attraction of two elongated orbits can be
approximated by considering each real oval precessing orbit as a
``needle'', which coincides with a major oval axis; the linear density
of a needle is $\rho_l^{(E)} = 1/v_r(E)=1/\sqrt{2E-2\Phi_0(r)}$, $E$
is the star energy, $v_r$ is the star radial velocity, $M_1(E) =
\int\limits_0^a \rho_l^{(E)}(r)dr$ is the half of needle mass, $2a$ is
the needle length, and
$$
\Omega'_{pr}(E,L)=\left.\frac{\partial \Omega_{pr}}{\partial
  L}\right|_{J_f} = \frac{\partial \Omega_{pr}}{\partial  L} + 
\Omega_{pr}\frac{\partial \Omega_{pr}}{\partial E}.
$$

\begin{figure}[t]
\begin{center}
\includegraphics[width=8cm]{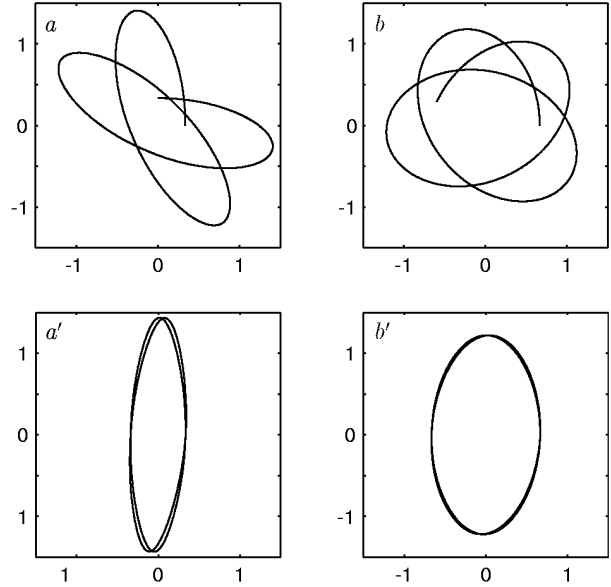}
\end{center}
\caption[]{Typical orbits of stars in the Shuster potentials 
for the models studied by Athanassoula and Sellwood (1986): {\it a} --
models with the smallest mean precession speeds and angular momenta;
{\it a}\,$'$ -- the same as {\it a} but in the reference frame rotating
with the precessing orbit; {\it b} --  majority of models; {\it b}\,$'$
-- the same as {\it b} but in the reference frame rotating with the
precessing orbit.} 
\label{fig1}
\end{figure}

By using the equation (\ref{i20}) Polyachenko (1992) had computed the
``abnormally'' low-frequency bar-modes, which were found by
Athanassoula and Sellwood (1986) in their $N$-body study of the linear
stability of some exact phase models of stellar disks. Those
frequencies were ``abnormally'' low compared to the frequencies of
``standard'' fast bars, obtained by them for the majority of the
models studied. Actually, the pattern speeds of the low-frequency
modes are approximately equal to mean orbit precession speeds in a
central disk region. Hence it follows that the instability of 
elongated orbits occurs. Figs.\,\ref{fig1}a,\,a$'$ show a typical
orbit participating in the slow bar-mode instability; this orbit
corresponds to mean values of energy and angular momentum of stars
over the region of the bar location. As one can see, the orbit is
strongly elongated; this fact justifies the use of the equation
(\ref{i20}) with the kernel (\ref{k21}). The computed growth rates
(Polyachenko 1992) are in good agreement with those obtained in the
paper by Athanassoula and Sellwood (1986). On the other side,
Figs.\,{fig1}b,\,b$'$ show the analogous typical orbit for the models
in which only the fast bar mode developed. This orbit is substantially
more round, than the orbit in Figs.\,\ref{fig1}a,\,a$'$. So, for
obtaining the low-frequency eigen modes in such models, the use of the
general integral equations (\ref{i8}) or (\ref{i10}) seems to be more
adequate. However, we think that even the integral equation
(\ref{i20}), being certainly a rough approximation for these models,
can provide a satisfactory numerical agreement with $N$-body results
of Athanassoula and Sellwood (1986). We plan to study all these
problems elsewhere.

\section{Bar-mode in the model two-component disk}

Let us represent the dispersion relation that can be derived from (\ref{i20}) for 
``one-component'' system with the equilibrium distribution function
$f_0 = A\delta(E - E_0^{(1)}) \delta(L - L_0^{(1)})$, in the following form:
\begin{equation}
1+\frac{g_1}{(\Omega_p - \Omega_1)^2} = 0,
\label{de22}
\end{equation}
where $\Omega_1 \equiv \Omega_{pr}(E_0^{(1)}, L_0^{(1)})$ and 
$g_1 \propto \Omega'_{pr}(E_0^{(1)}, L_0^{(1)})$ denotes the
corresponding coefficient; its explicit expression can be easily
obtained from (\ref{i20}) after substituting a given
$\delta$--distribution function. In (\ref{de22}), the instability corresponds
to $g_1 < 0$. 

Below we consider the case of the two-component system, with the
distribution function $f_0 = A\delta(E - E_0^{(1)}) \delta(L -
L_0^{(1)}) +  B\delta(E - E_0^{(2)}) \delta(L - L_0^{(2)})$. Then some
simple manipulations with the equation that is obtained after
substituting this distribution function into (\ref{i20}) lead to the
dispersion relation: 
\begin{equation}
1+\frac{g_1}{(\Omega_p - \Omega_1)^2} +\frac{g_2}{(\Omega_p -
  \Omega_2)^2} + \alpha\frac{g_1g_2}{(\Omega_p - \Omega_1)^2(\Omega_p -
  \Omega_2)^2} = 0,
\label{de23}
\end{equation}
where $g_2 \propto \Omega'_{pr}(E_0^{(2)}, L_0^{(2)})$ is analogous to
the coefficient $g_2$ for the first component introduced earlier,
$\Omega_2 \equiv \Omega_{pr}(E_0^{(2)}, L_0^{(2)})$, and the
designations are used: 
\begin{equation}
\alpha = 1-\frac{I_0^2(E_0^{(1)},E_0^{(2)})}{I_0(E_0^{(1)},E_0^{(1)})
  I_0(E_0^{(2)},E_0^{(2)})} 
\label{r24}
\end{equation}
\begin{equation}
I_0(E_0^{(i)},E_0^{(j)}) \equiv \int\limits_0^{a_i}\int\limits_0^{a_j} dx\,dy\,
\rho^{(E_i)}(x) \rho^{(E_j)}(y) J_m(x,y),
\label{r25}
\end{equation}

It turns out that the possible locations of $\mathrm{Re}\,\Omega_p$
for the unstable roots ($\gamma = \mathrm{Im}\,\Omega_p > 0$)
essentially depends on the signs of the coefficients  $g_1$ and $g_2$. 

Let us prove first of all that for positive $g_1$ and $g_2$ the
inequalities  $\Omega_1 < \mathrm{Re}\,\Omega_p <\Omega_2$ occur. To
do this, it is sufficient to calculate  the imaginary part of the left
side of the dispersion relation (23). We have 
$$
A_{1,2} \equiv \mathrm{Im}\,\frac{g_{1,2}}{(\Omega_p - \Omega_{1,2})^2} =
-\frac{2g_{1,2} \gamma \Delta_{1,2}}{(\Delta_{1,2}^2 + \gamma^2)^2}, \quad
\Delta_{1,2} \equiv \mathrm{Re}\,\Omega_p - \Omega_{1,2};
$$
$$
B \equiv \mathrm{Im}\,\frac{\alpha g_1g_2}{(\Omega_p -
  \Omega_1)^2(\Omega_p - \Omega_2)^2 }= -\frac{2g_1g_2\alpha
    \gamma}{(\Delta_1^2 + \gamma^2)^2(\Delta_2^2 + \gamma^2)^2}
[(\Delta_2^2+\gamma^2)\Delta_1 + (\Delta_1^2+\gamma^2)\Delta_2], 
$$
Assuming $\mathrm{Re}\,\Omega_p > \Omega_2$ we have $\Delta_1 > 0,
\Delta_2 > 0$. So in this case the imaginary part of the left side of
the dispersion relation (23) would be negative: 
$$
A_1 + A_2 +B < 0,
$$
if one takes into account that $\gamma > 0$ for the unstable roots of
interest and, besides, $\alpha > 0$ as it follows from the
Cauchi--Bunyakovsky inequality (positive definiteness of the weight
function $J_2(x, y)$ was proved in the preceding section). 

Similarly, for $\mathrm{Re}\,\Omega_p < \Omega_1$ all inequalities are 
reversed (of course, except for $\alpha > 0$): $\Delta_1 < 0, \Delta_2
< 0$, so 
$$
A_1 + A_2 + B > 0.
$$

\begin{figure}
\begin{center}
\includegraphics[width=10cm]{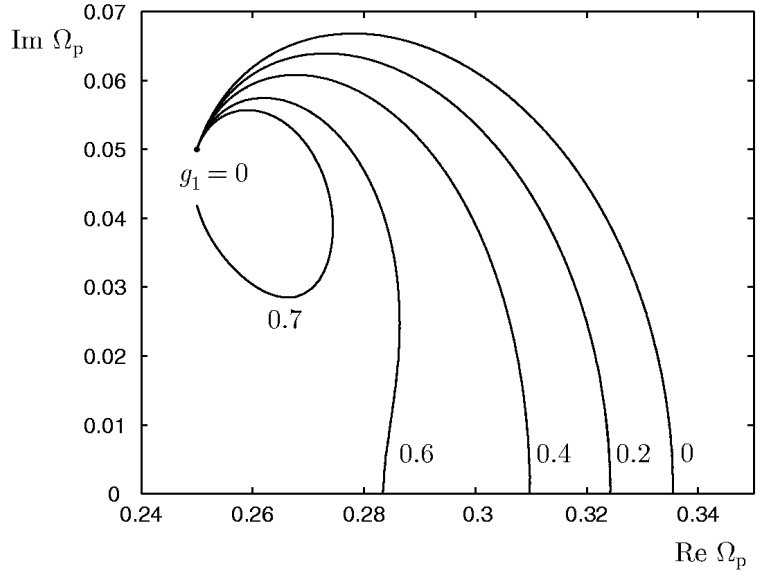}
\end{center}
\caption[]{The trajectories of the unstable root in the two-component
  model with opposite signs of $g_1<0$ and $g_2>0$, for a number of values
  of the coefficient $\alpha =0$, $0.2$, ... $0.7$. The absolute value of $g_1$ increases
  along the trajectories from the starting point $g_1 = 0$.}
\label{fig2}
\end{figure}

A completely different type of situation occurs for the case when the
signs of $g_1$ and $g_2$ are opposite, i.e. the disk contains the
orbits with a ``donkey'' behavior;  for definiteness, we assume that
$g_2 > 0$, $g_1 < 0$.  Fig.\,\ref{fig2} show the trajectories of motion of the
unstable root on the complex plane $\Omega_p$, for a fixed $g_2 =
0.05^2 > 0$ ($\Omega_1 = 0$ and $\Omega_2 = 0.25$ are also fixed) and
negative $g_1$ that vary along each trajectory from $g_1 = 0$ to some
$(g_1)_{\min}$, at  which this root become stable too. Different
trajectories correspond to different values  of the parameter
$\alpha$. The most important fact following from these calculations is
that the excess of $\mathrm{Re}\,\Omega_p$ over $\Omega_2$ can be
quite significant (about 1.5 times in a given example).

\section{The patterns of orbit precessions in some typical potentials}

\begin{figure}[t]
\begin{center}
\includegraphics[width=11cm]{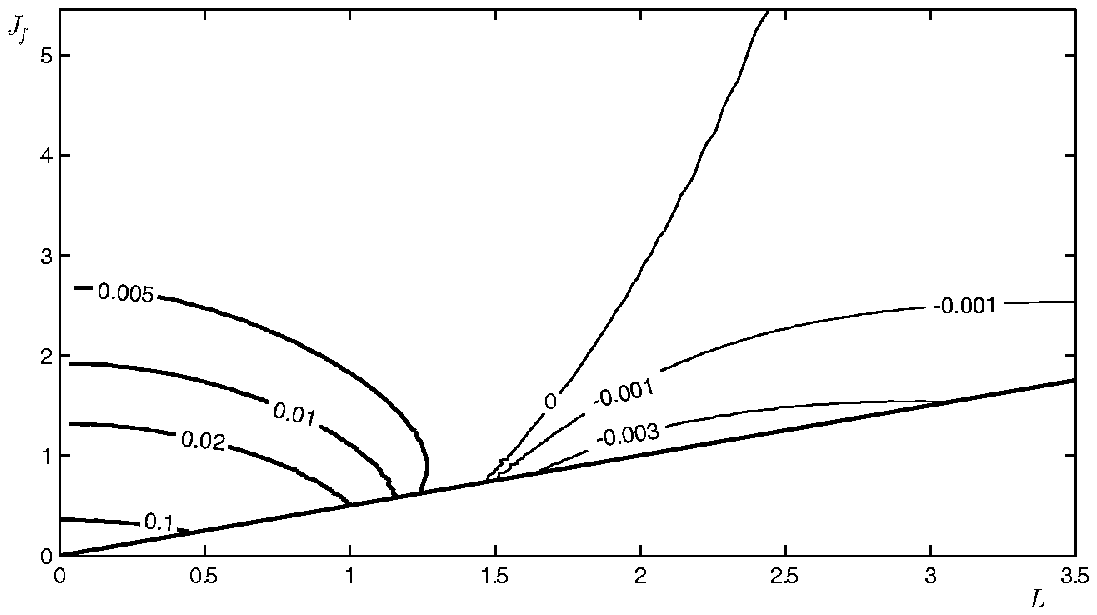}
\includegraphics[width=14cm]{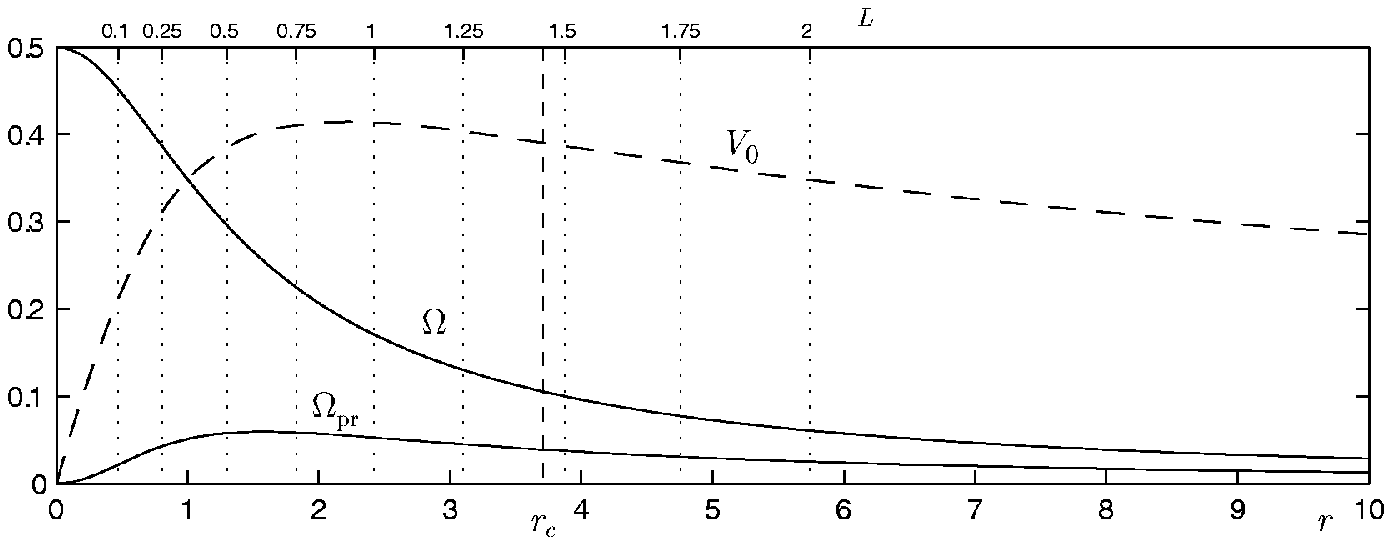}
\end{center}
\caption[]{{\it a} -- the pattern of orbit precessions for the
  isochrone model at the Lynden-Bell ($J_f$, $L$)-plane. The strait
  line $J_f = L/2$ corresponds tocircular orbits. The curves are the
  isolines for the Lynden-Bell derivative $\Omega'_{pr}$. {\it b} --
  the angular disk velocity $\Omega(r)$, the rotation curve $V_0(r)$
  and the circular orbit precession speed $\Omega_{pr}(r)$ for the
  isochrone model.}
\label{fig3}
\end{figure}

\begin{figure}[t]
\begin{center}
\includegraphics[width=11cm]{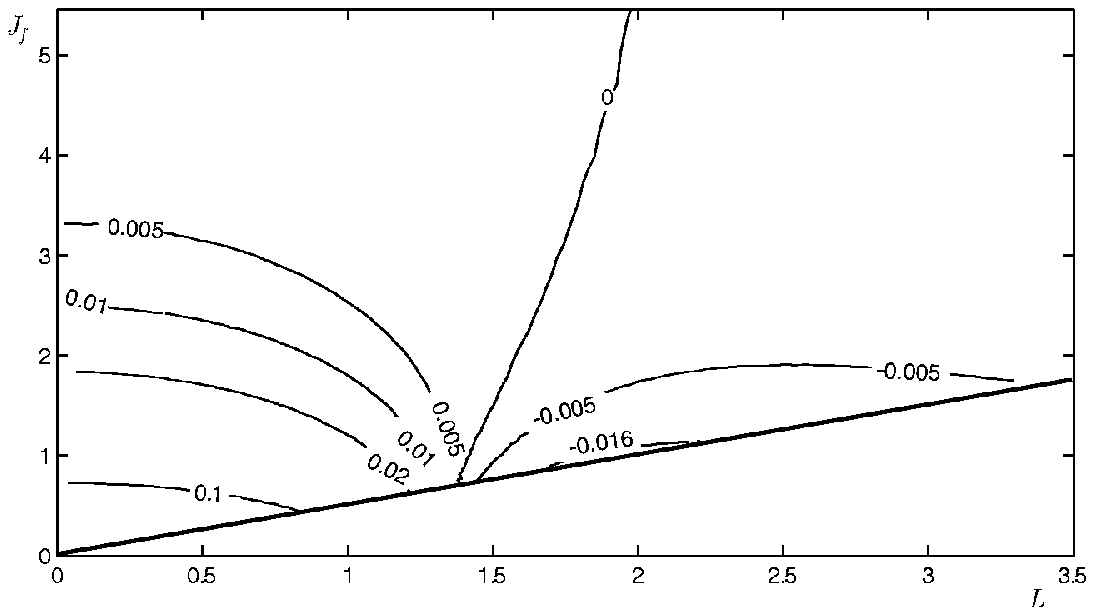}
\includegraphics[width=14cm]{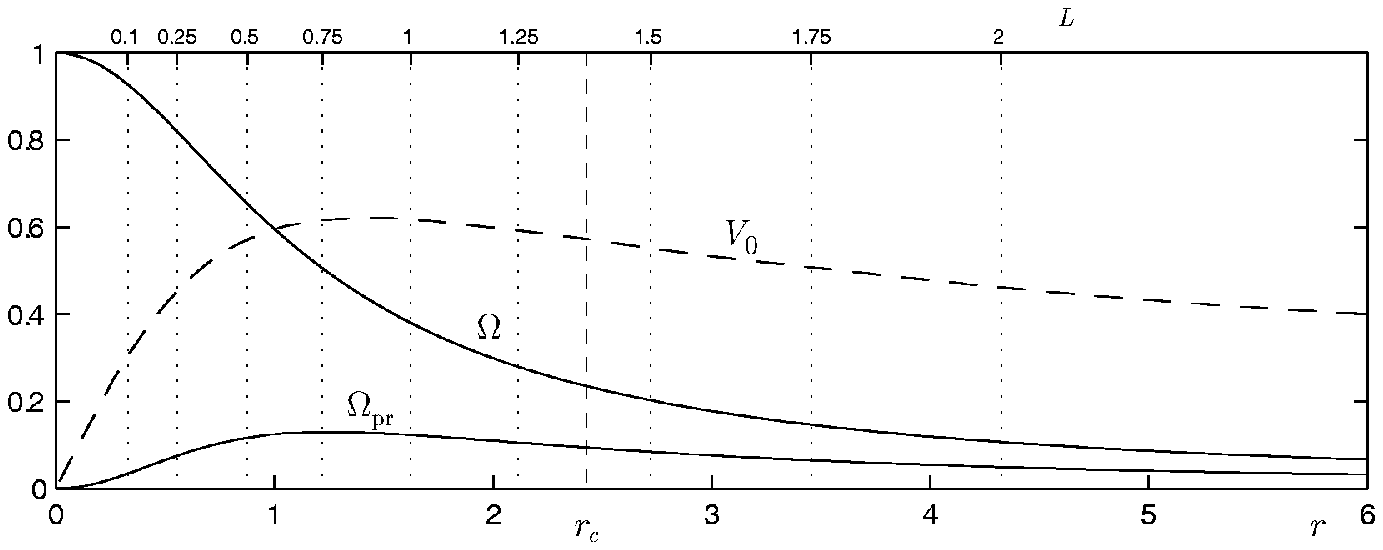}
\end{center}
\caption[]{The same as in the Fig.\,\ref{fig3} for the Shuster potential.}
\label{fig4}
\end{figure}

\begin{figure}[t]
\begin{center}
\includegraphics[width=11cm]{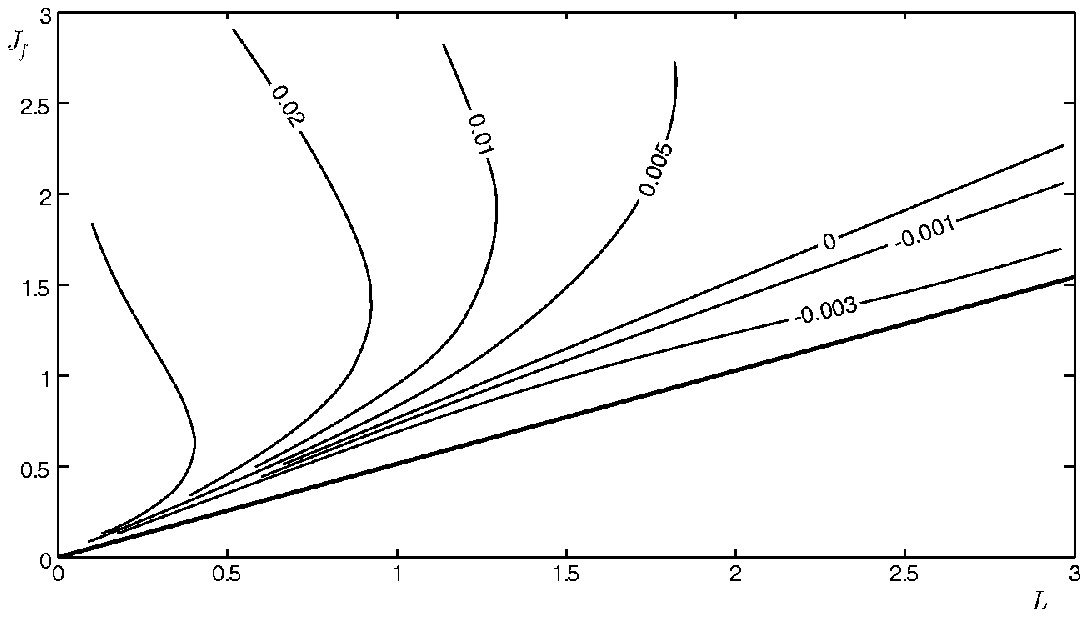}
\includegraphics[width=14cm]{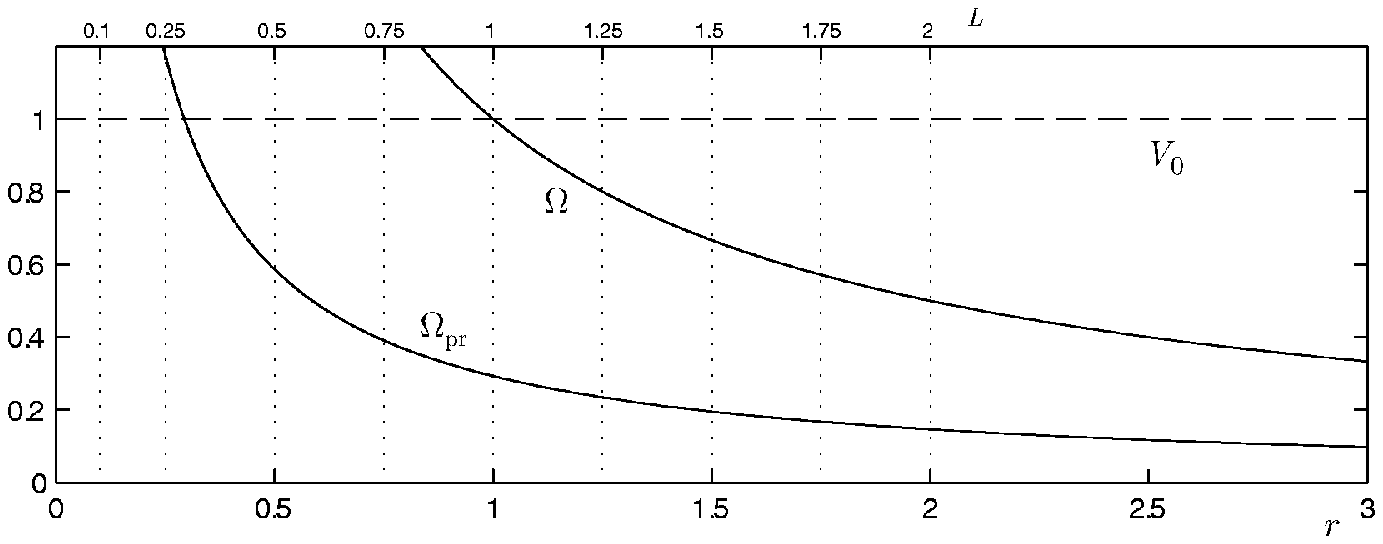}
\end{center}
\caption[]{The same as in the Fig.\,\ref{fig3} for the logarithmic potential.}
\label{fig5}
\end{figure}

\begin{figure}[t]
\begin{center}
\includegraphics[width=11cm]{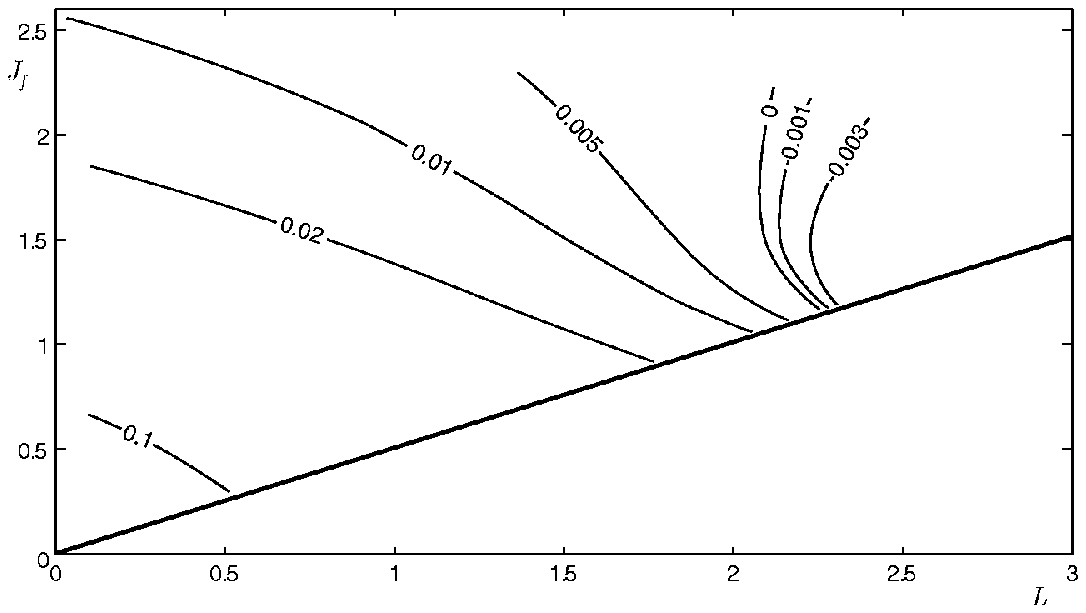}
\includegraphics[width=14cm]{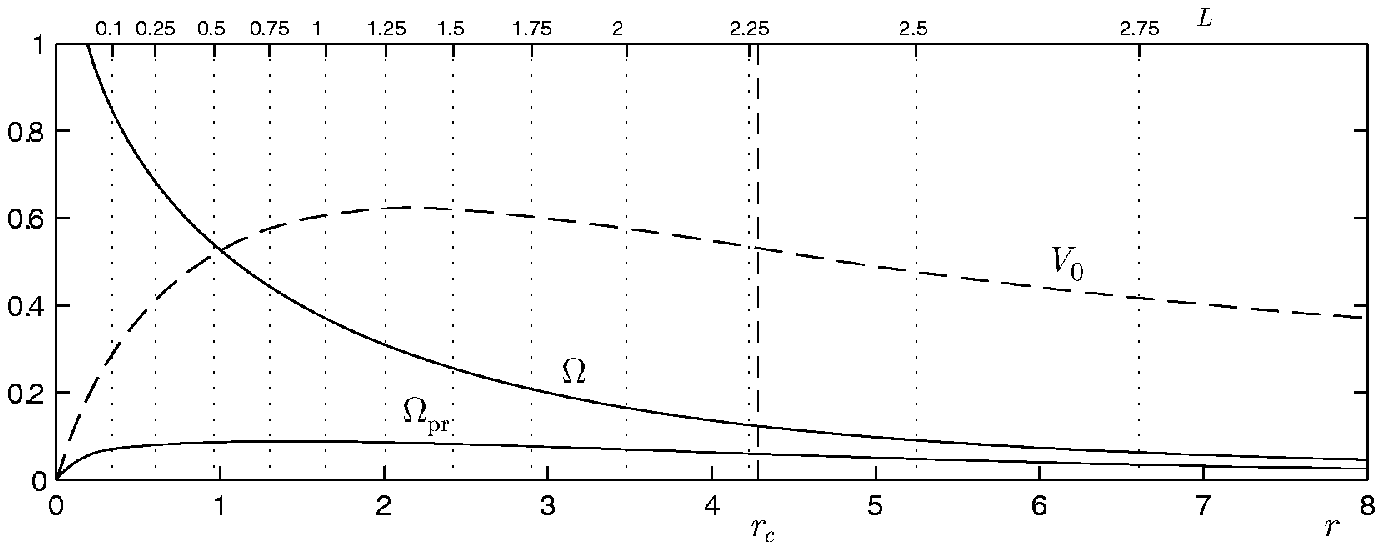}
\end{center}
\caption[]{The same as in the Fig.\,\ref{fig3} for the potential of
  the exponential disk.}
\label{fig6}
\end{figure}

Fig.\,\ref{fig3}a--\ref{fig6}a show, on the Lynden-Bell plane $(J_f,
L)$, the constant 
value curves for the  derivative $\Omega'_{pr} \equiv (\partial
\Omega_{pr}/\partial L)_{J_f}$, for a number of the commonly occuring
potentials $\Phi_0(r)$. In the parallel
Figs.\,\ref{fig3}b--\ref{fig6}b, we give, for  
the same potentials, the angular velocities of stars at the circular
orbits $\Omega (r)$,  tthe rotation curves $V_0(r) = r\Omega(r)$   and
the precession speeds of the nealy-circular orbits  $\Omega_{pr}(r) =
\Omega (r) - \kappa(r)/2$ ($\kappa(r)$ is the epicyclic frequency).

The first pair of these figures (Figs.\,\ref{fig3}a,\,b) correspond
to the isochrone potential  
$$
\Phi_0(r) = -\frac1{1+\sqrt{1+r^2}}.
$$
This case was earlier considered by Lynden-Bell (1979). For this
potential, the frequencies  $\Omega_1(J_f, L)$, $\Omega_2(J_f, L)$ as
well as the quantities of interest  $\Omega_{pr}(J_f, L) =
\Omega_2(J_f, L) - \Omega_1(J_f, L)/2$ and $\Omega'_{pr}(J_f, L)$ can
be  obtained analytically. We used the example of the isochrone
potential as a test model for  our general scheme of computation of
$\Omega_{pr}$ and $\Omega'_{pr}$ for the arbitrary  potential $\Phi_0(r)$. 

Apart from the isochrone model, we carried out the computations for
the Shuster potential 
$$
\Phi_0(r) = -\frac1{\sqrt{1+r^2}}
$$
(Figs.\,\ref{fig4}a,\,b),the logarithmic potential $\Phi_0(r) =\ln r$,
corresponding to the flat  rotation curve $V_0 = \mathrm{const}$
(Figs.\,\ref{fig5}a,\,b), and for the potential of the exponential disk, 
$$
\Phi_0(r) = rI_1(r/2)K_0(r/2),
$$
where $I_1$ and $K_0$ are the corresponding Bessel functions
(Figs.\,\ref{fig6}a,\,b). Qualitatively,
Figs.\,\ref{fig3}a,\,\ref{fig4}a,\,\ref{fig6}a  and
\ref{fig3}b,\,\ref{fig4}b,\,\ref{fig6}b  are similar, 
but they differ greatly from Figs.\,\ref{fig5}a,\,b for the  logarithmic
potential (and it is not unreasonable). In our opinion, the most
interesting information that one can extract from
Figs.\,\ref{fig3}a,\,\ref{fig4}a,\,\ref{fig6}a are the
critical values of the angular momentum ($L_c$) that separate the
regions of ``attracting'' (when $L < L_c$) and ``repelling'' (when $L
> L_c$) near-circular orbits.The critical radii ($r_c$) corresponding
to these values of $L_c$ are shown in 
\ref{fig3}b,\,\ref{fig4}b,\,\ref{fig6}b.
It is natural
to take  $r_c$ as an estimate for a bar length $l_b$ that forms as a
result of the bar-instability  under consideration (a least at the
linear stage). As one can see, $r_c$ is always more than  the radii
corresponding to maxima of the rotation curve $V_0(r)$ or the function
$\Omega_{pr}(r)$. Under the condition commonly used that a fast bar
ends near the corotation, we obtain from $l_b \sim r_c$ such an
estimate of the bar pattern speed, for the Shuster potential:
$\Omega_p \approx 0.23$ (see Fig.\,\ref{fig4}b). The corresponding eigen
frequency $\omega = 2\Omega_p \approx 0.46$ is typical for the
majority of models studied by Athanassoula and Sellwood (1986),  
just for the Shuster potential. Note that these bars are traditionally
considered as ``fast'' ones, keeping in mind their non-Lynden-Bell
formation mechanism. We see, however, that the fast bars can likely
be formed by the same mechanism as for the slow bars. Let us point out
some other interesting regularities that are common for the patterns
of the orbit precessions in potentials $\Phi_0(r)$ of a type
corresponding to Figs.\,\ref{fig3}a,\,\ref{fig4}a,\,\ref{fig6}a: 

1. The precession speed $(\Omega_{pr}^{\max})$ corresponding to the
   maximum of the curve $\Omega_{pr}(r)$ (at the circular orbits) is
   the absolute maximum for the precession speeds $\Omega_{pr}(J_f,L)$
   of arbitrary orbits; $(\Omega_{pr}^{\max})$ for three models are
   given in the first line of the table.

\vspace{12pt}
\centerline{
\begin{tabular}{|c|c|c|c|}
\hline
  &Isochrone&Shuster&$\exp$ disk \\
\hline
$(\Omega_{pr}^{\max})$&$ 0.058$&$0.13$&$0.087$\\
\hline
$r_c$&$3.7$&$2.4$&$4.3$\\
\hline
$({\Omega'}_{pr}^{\max})$&$0.3 \div 0.35$ & $0.4\div0.45$ & $0.35$\\
\hline
$({\Omega'}_{pr}^{\min})$&$-0.0105$&$-0.021$&$-0.025$\\
\hline
\end{tabular}
}

\vspace{12pt}
2. $({\Omega'}_{pr}^{\max})$ are approached at the circular orbits in
   the disk center ($J_f \to J_r \to 0$, $L\to 0$); the values of
   $({\Omega'}_{pr}^{\max})$ for three models are given in the third
   line of the table.

3. $({\Omega'}_{pr}^{\min})$ are approached in some points at the
   circular orbits; the values of  $({\Omega'}_{pr}^{\min})$ are given
   in the fourth line of the table.

The pattern of the orbit precession in the logarithmic potential
(typical for the major parts of many spiral galaxies) is entirely
different from the other cases considered (see Fig.\,\ref{fig5}a). The
most interesting fact here is that $\Omega'_{pr} < 0$ only within a
rather narrow sector of the ($J_f$, $L$)-plane, adjacent to the line
of circular orbits $J_f = L/2$. 

The massives of the values of functions $\Omega_{pr}(J_f, L)$,
$\Omega'_{pr}(J_f, L)$ obtained above will be used later under
studying the integral equations (\ref{i8}), (\ref{i10}) for deriving
the eigen bar-modes (first of all, the stellar models in the Shuster
potential).

\section{Conclusion}
Let us formulate and discuss some conclusions from the theory above.  

1. We advanced a number of arguments in favor of universality of the
   mechanism that may be responsible for formation of both the slow
   and fast bars. The essence of this mechanism is naturally
   formulated on the basis of representing a stellar disk as a set of
   precessing orbits. Such a concept is adequate to the problem under
   consideration since the bar pattern speeds (including ``fast bars'')
   are significantly less than the characteristic frequencies of
   oscillations of individual stars ($\Omega_1$ and $\Omega_2$), but
   they are just of order of the orbit precession speeds
   ($\Omega_{pr}$). In such a disk, we seek the unstable normal modes
   (first of all the bar-mode) as the density wave of precessing
   orbits that runs with some speed $\Omega_p$ without any
   deformations despite the fact that different orbits precess with
   different speeds (analogously to differentiability of a disk rotation
   when the usual concept of a galactic disk as a set of individual
   stars is used).  

The unstable bar-mode forms if a central region of a disk (a location
of a future bar) contains a sufficiently massive group of ``attracting''
orbits that satisfy the Lynden-Bell condition $\Omega'_{pr} > 0$, and,
besides, the precession speed dispersion of these active orbits are
not too large (otherwise, the orbits will run away from the region of
perturbation under the influence of the ``thermal'' motion). The last
condition is natural for the Jeans nature of the instability under
consideration. The exact criteria of instability can be obtained by
solution of the basic equations derived in Sec. 2; for some simple
cases, the corresponding dispersion relations are given in an explicit
form. The pattern speed of a bar $\Omega_p$ depends significantly on
the extent to which the ``repelling'' orbits with $\Omega'_{pr} < 0$ (the
orbits with a ``donkey'' behavior) take part in the bar-formation
process. If such orbits are hardly dragged in the bar-formation
process, then $\Omega_p \sim \bar\Omega_{pr}$; just such bars would
naturally be named as the ``slow'' bars. But if the role of the
``repelling'' orbits is essential, we can obtain the ``fast'' bars with
$\Omega_p$ that significantly exceeds $\bar\Omega_{pr} $(just the
same, $\Omega_p$ should be of order of $\bar\Omega_{pr}$ as before --- if
we want to remain within the framework of our theory). In Sec. 3, such
a possibility is demonstrated on the simplest example of a
two-component disk model. Thus, from the point of view under
consideration, distinctions between the slow and fast bars are mainly
quantitative, but they do not differ fundamentally from each other:
both bars form under the action of the same physical mechanism. It is
worth noting that the Jeans mechanism (including one under
consideration) is always best natural and suited to the gravitational
problems. 

2. So far the theory above was confirmed only by the calculations of
   the lowest-frequency modes of the disk models in the Shuster
   potential when the results are compared with the $N$-body results by
   Athanassoula \& Sellwood (1986). These modes obviously correspond to
   the slow bars: for them, $\Omega_p \sim \bar\Omega_{pr}$. The most
   interesting prediction of the theory (its validity for the fast
   bars) would be tested if we shall be able to prove that the eigen
   frequencies of the modes computed from our integral equations (see
   Sec. 2) coincide with the frequencies of the fast bars derived
   from the $N$-body simulations (in particular, with the majority of
   bar-modes from the paper by Athanassoula \& Sellwood cited
   above. This problem will be the subject of study in the immediate
   future. In the present paper, we restricted ourselves only to some
   positive facts that result from the general analysis of orbit
   precessions for a number of potentials (Sec. 4). These facts
   correlate with a possibility of the fast bar formation by alignment
   of ``attracting'' orbits. First of all we noted that the maximum
   radius of nearly-circular orbits ($r_c$) with $\Omega'_{pr} > 0$
   ($r_c$ may be taken as a natural estimate for a bar end $l_b$) is
   significantly more than a size ($\sim r_m$) of the solid rotation
   region, for all the reasonable models. Moreover, if one takes (as
   it usually does) that $l_b \sim r_{CR}$ (where $r_{CR}$ is the
   corotation radius), then for $l_b \sim r_c$ one can obtain the
   estimate $\omega \approx 0.46$ for a typical eigen frequency
   $\omega$ of bar-modes in the Shuster potential; this estimate is in
   good agreement with the corresponding results of Athanassoula \&
   Sellwood (1986).

\end{document}